\documentclass[
    ,final            
  ]
  {aipproc}

\layoutstyle{6x9}

\usepackage{xspace}
\usepackage[tight]{subfigure}
\usepackage{graphicx} 

\makeatletter
\newcommand*{\SubLabels}[1]{%
  \label{fig:#1}%
  \begingroup
    \protected@edef\@currentlabel{%
      \csname thesub\@captype\endcsname
    }%
    \label{subfig:#1}%
  \endgroup
}
\makeatother

\newcommand{\pT}{\ensuremath{p_T}\xspace}
\newcommand{\vTwo}{\ensuremath{v_2}\xspace}

\newcommand{\piz}{\ensuremath{\pi^0}\xspace}
\newcommand{\RAA}{\ensuremath{R_{AA}}\xspace}

\newcommand{\dphi}{\ensuremath{\Delta\phi}\xspace}

\begin{document}


\title{High-\pT Particle Production in PHENIX}


\author{David L. Winter, for the PHENIX Collaboration}{
  address={Columbia University, NY, NY U.S.A.}
}

\begin{abstract}
It has been established that "hard probes", observables involving
high-momentum transfer, provide useful tools for studying the hot,
dense medium created in nucleus-nucleus collisions at RHIC.  The
nuclear modification factor, azimuthal correlations, direct photon
production, as well as the dependence of the nuclear modificaton
factor on centrality and angle with respect to the reaction plane are
critical for understanding the early dynamics of such heavy-ion
collisions.  We will review recent results from PHENIX for particle
production at high-\pT and discuss their implications.
\end{abstract}

\maketitle


\section{Introduction}

PHENIX is an ongoing experiment at the Relativistic Heavy-Ion Collider
(RHIC) at Brookhaven National Laboratory, dedicated to searching for
evidence of a phase transition from normal nuclear matter to the Quark
Gluon Plasma (QGP).  The QGP is a phase of matter consisting
deconfined quarks and gluons expected to be formed at high energy
densities (above $\approx 1$~GeV/fm${^3}$).  

High-\pT observables are a critical probe for understanding the
evolution of the collision region.  Specifically PHENIX excels in
measuring high-\pT neutral mesons and photons.  An important tool for
characterizing the production of high-\pT photons and hadrons, and for
studying medium-induced effects on such particle production, is the
nuclear modification factor \RAA:
\begin{equation}
R_{AA}(\pT) = \frac{d^2N_{AA}/dyd\pT}{\langle T_{AA}\rangle d^2N_{pp}/dyd\pT}
\end{equation}
where $\langle T_{AA}\rangle$ is the nuclear overlap function for the
given centrality of the collision.  The denominator represents the
expected yield supposing the A+A collision is a super-position of
nucleon-nucleon collisions.  Therefore, for the high-\pT region
($>2$~GeV/c) where the physics is dominated by hard-scattering of
partons, any deviation from unity is expected to arise from medium
effects. 

\section{PHENIX High-\pT Measurement}

The PHENIX detector is described in~\cite{ref:phenix}.  High-\pT
neutral mesons and direct photons are measured using the
Electromagnetic Calorimeter (EMCal)~\cite{ref:emcal}. The EMCal
consists of 8 sectors covering $\pi$ radians in azimuth and
$|\eta|<0.35$ in pseudorapidity.  For measurements made with respect
to the reaction plane, the reaction plane orientation is determined on
an event-by-event basis using the Beam-Beam-Counters
(BBCs)~\cite{ref:reacpl,ref:Cole}.

\begin{figure}
\includegraphics*[height=0.2\textheight]{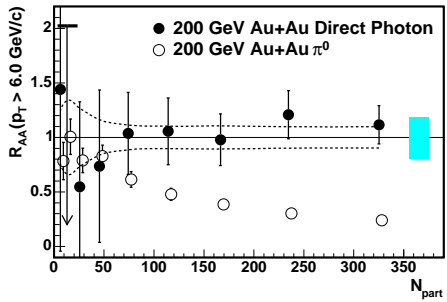}
\caption{\RAA (integrated above $\pT>6$~GeV/c) as a function of
centrality ($N_{part}$) for direct photons and \piz{}s in 200 GeV
Au+Au collisions. The band shows the uncertainty in the pQCD
normalization.}
\label{fig:pi0_photon_raa}
\end{figure}

An important baseline measurement that established the validity of
perturbative QCD as it applies to heavy-ion collisions, as well as
provide indisputable evidence of hard-scattering occuring at RHIC
collisions, is the \RAA of direct photons~\cite{ref:Adler:2005ig}.
Figure~\ref{fig:pi0_photon_raa} shows the direct photon \RAA as a
function of the number of participants, $N_{part}$. Also shown is the
\RAA for \piz{}s.  The comparison is striking: the photons have an
\RAA consistent with unity, while the
\piz{}s exhibit suppression.  Since the photons will not be subject to
final state effects, this represents clear evidence that the
suppression is due to hadronic medium effects.  We now know that this
has to be due to final state effects, as the suppression is absent in
d+Au collisions~\cite{ref:Adler:2003}.

\begin{figure}
\begin{minipage}{\textwidth}
\includegraphics[width=0.49\textwidth]{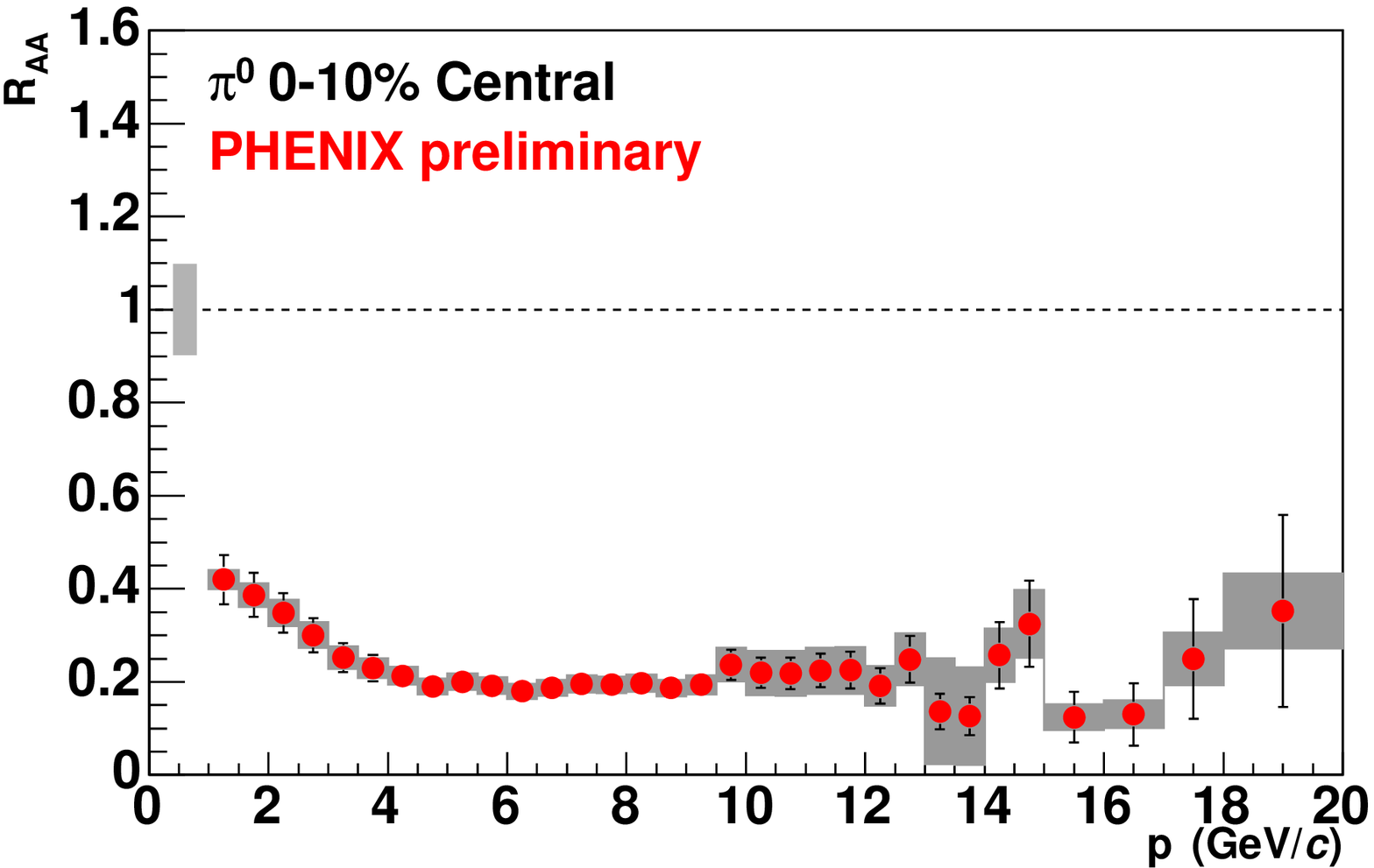}
\includegraphics[width=0.49\textwidth]{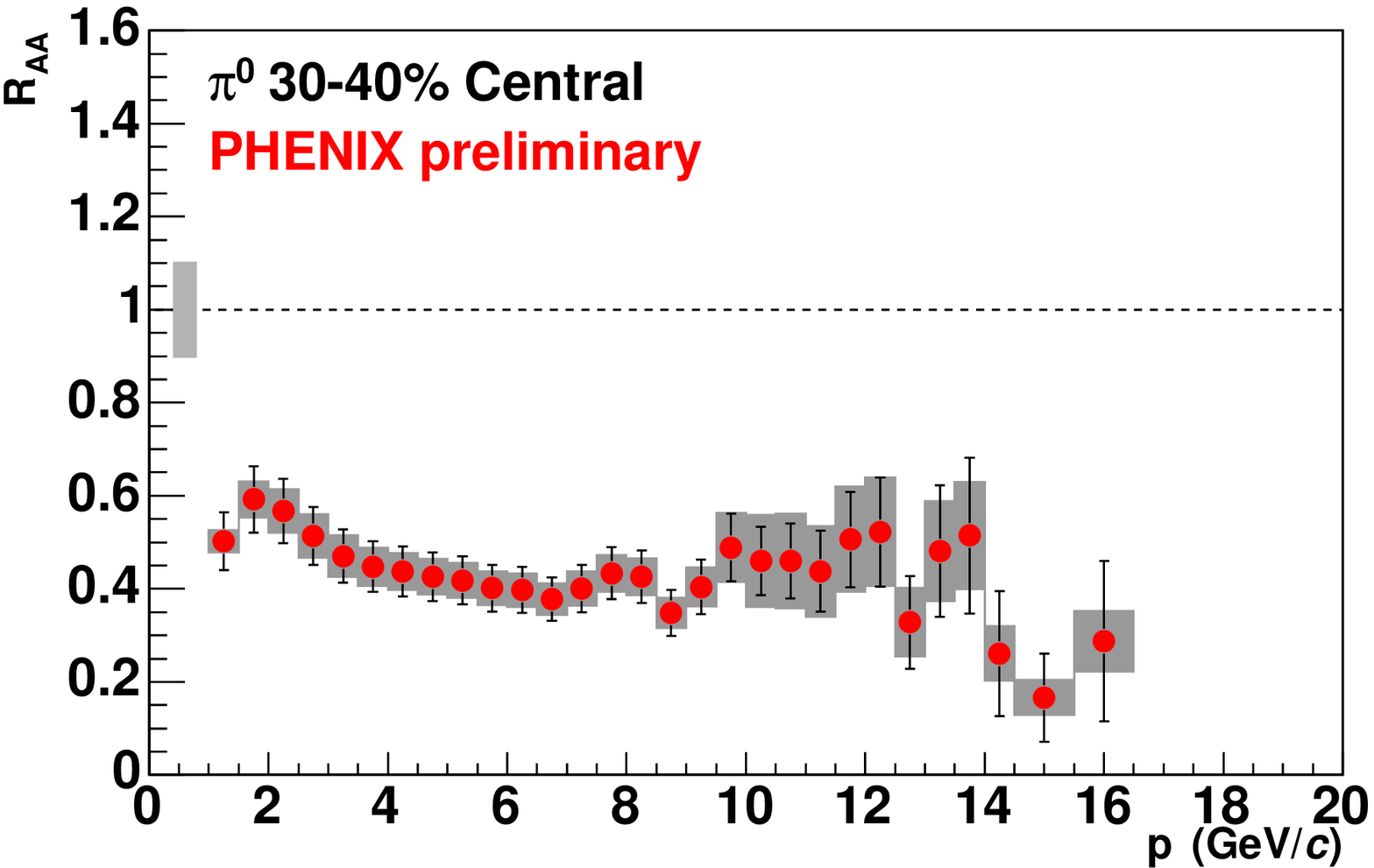}
\end{minipage}
\caption{Inclusive \piz $\RAA(\pT)$ 200 GeV/c Au+Au collisions. The
boxes are the \pT-dependent systematic errors, and the systematic
error on the normalization is shown as the box around unity.}
\label{fig:pi0RAA}
\end{figure}

Preliminary inclusive \piz \RAA is shown in Figure~\ref{fig:pi0RAA},
for two sample centrality classes of $\sqrt{s_{NN}}=200$ GeV Au+Au
collisions from RHIC Run 4. We see that in central collisions the
previously observed suppression exists up to $\pT \approx 20$~GeV/c.
The GLV model~\cite{ref:GLV} describes the suppression well, and
implies the data are consistent with $dN_g/dy \approx 1100$ and an
energy density $\approx 15$~GeV/fm$^3$.

\begin{figure}
\includegraphics*[width=\textwidth]{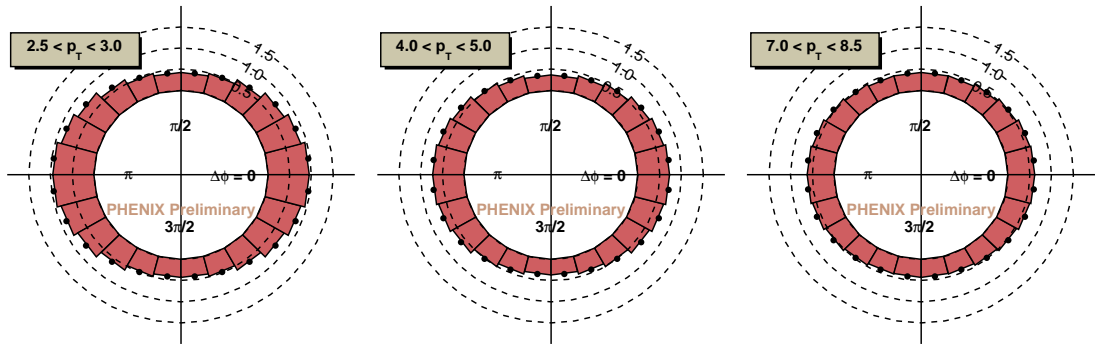} 
\caption{Polar plots of \piz \RAA(\pT,\dphi) for 30-40\% central 200 GeV Au+Au
   collisions, for three selected \pT ranges.  The measured $0-\pi/2$
   rad data are folded to display the full $2\pi$ range. Errors shown
   are statistical only.}
\label{fig:raa_dphi}
\end{figure}

In order to help constrain energy loss models that attempt to
describe hadronic suppression, an analysis of \RAA as a function of
angle of emission with respect to the reaction has been
proposed~\cite{ref:Cole}.  Preliminary results of measurements of this
observable are shown in Figure~\ref{fig:raa_dphi}.  We note that
although $\RAA(\dphi)$ does not contain any new information that is
not already available in a separate measurement of \RAA and the
elliptic flow parameter \vTwo, this is the first type of measurement
that combines those two observables in a single measurement.  The
angular variation of \RAA for a given \pT range can be compared to the
corresponding data points in Figure~\ref{fig:pi0RAA}.

There are other high-\pT observables, the description of which is
beyond the scope of this paper.  Azimuthal
anisotropy~\cite{Winter:2005nw} and jet structure via two-particle
correlations~\cite{ref:Stankus} are two examples.

\section{Summary}

PHENIX measures a broad range of high-\pT observables targeted at
understanding the transition from normal nuclear matter to the QGP.
Presented here are highlights of measurements of \RAA for photons and
\piz{}s observed in $\sqrt{s_{NN}}=200$ GeV Au+Au collisions.  These
data show clear evidence for strong hadronic suppression in central
collisions, which is attributed to final state effects.  Furthermore,
the central data are consistent with $dN_g/dy \approx 1100$ and large
energy density.  Also presented is the PHENIX measurement of \RAA with
respect to the reaction plane, which will be an effective to tool to
allow more detailed study of the models used to describe energy loss
at high-\pT.





\bibliographystyle{aipproc}   

\bibliography{CIPANP06_Winter_Proceedings}

\IfFileExists{\jobname.bbl}{}
 {\typeout{}
  \typeout{******************************************}
  \typeout{** Please run "bibtex \jobname" to optain}
  \typeout{** the bibliography and then re-run LaTeX}
  \typeout{** twice to fix the references!}
  \typeout{******************************************}
  \typeout{}
 }

\end{document}